\documentclass[review]{elsarticle}
\usepackage[hypertex]{hyperref}
\usepackage{lineno,hyperref}
\usepackage{color}
\modulolinenumbers[5]
\journal{European Journal of {Operational} Research}
\renewcommand{\a}{\alpha}
\renewcommand{\b}{\beta}
\newcommand{\bea}{\begin{eqnarray}}
\newcommand{\eea}{\end{eqnarray}}
\newcommand{\f}[2]{\frac{#1}{#2}}
\newcommand{\eq}{&=&}
\newcommand{\nn}{\nonumber \\ }
\newcommand{\ve}{\varepsilon}
\renewcommand{\d}{\delta}
\newcommand{\area}{\int_{-\infty}^\infty }

\renewcommand{\l}{\lambda}
\newcommand{\sref}[1]{eq. (\ref{#1})}
\newcommand{\p}{\partial}
\newcommand{\pp}[2]{\f{\p #1}{\p #2}}
\newcommand{\s}{\sigma}

\begin{document}

\begin{frontmatter}

\title{Self-Averaging Property of Minimal Investment Risk of Mean-Variance Model}

\author{Takashi Shinzato}
\address{{Department of Management Science and Engineering, Graduate School of Systems Science and Technology, Akita 
 Prefectural University,\\
shinzato@akita-pu.ac.jp}}
\begin{abstract}
In portfolio optimization problems, 
the minimum expected investment risk is not always smaller than the expected minimal investment risk.
That is, using a well-known approach from operations research, it is possible to derive a 
strategy that minimizes the expected investment risk, but this strategy 
does not always result in the best rate of 
return on assets. 
Prior to making investment decisions, it is important to an investor to know the potential minimal investment risk (or the
expected minimal investment risk) and to determine the strategy that will maximize the return on assets.
We {use} the self-averaging property to analyze the potential minimal investment risk and the concentrated 
investment level for the strategy that gives the best rate 
of return.
We {compare} the results from our method with the
results obtained by the operations research approach and with those obtained by a numerical simulation using the optimal 
portfolio. 
The results of our method and the numerical simulation {are} in agreement, but they {differ} from that of the operations research approach.
\end{abstract}

\end{frontmatter}
\linenumbers

\section{Introduction\label{sec1}}
{Investment is one }of the most common economic activities, and it is defined as an activity in which it is expected that future remuneration will more than repay the cost\cite{Dixit,Luenberger,Markowitz}. The uncertainty involved in investments cannot be removed, and in general, a greater risk corresponds to a greater expected return. 
In this paper, we consider the portfolio optimization problem, which is a mathematical formulation of risk management for investments. 
The portfolio optimization problem is based on the framework of risk diversification management, which was introduced by Markowitz in $1952$; 
it is the topic of some of the most important and most active 
{research} in mathematical finance, 
and various models have been proposed\cite{Konno,Luenberger,Markowitz,Rockafellar}.
For instance, 
Markowitz proposed a rule 
for investing in several securities 
in order to diversify.  For example, this rule states that
when the expected return and the invested assets are constant, 
the best strategy minimizes the variance of the return 
(investment risk). Markowitz also analytically 
derived the investment strategy which minimizes investment risk. 
Konno and Yamazaki 
proposed the mean-absolute deviation model, whose risk function is not defined by the variance of the  
return but as the sum of the absolute error in each 
period; it has also been shown that the optimal solutions of the mean-variance model 
and that of the mean-absolute deviation model are in agreement\cite{Konno}.
Rockafellar and Uryasev proposed an
expected shortfall model that is based on an index that measures the risk of not being less than a chosen confidence level; 
this model considers the downside risk of stochastically fluctuating gross earnings\cite{Rockafellar}.

In recent decades, the portfolio optimization problem has been studied using analytical approaches that were developed in cross-disciplinary fields other than operations research\cite{Ciliberti,Pafka,Shinzato-Yasuda,Wakai}. 
Ciliberti and M$\acute{\rm e}$zard used the replica analysis method developed in spin glass theory to analyze the typical behaviors of the
risk functions of the mean-absolute deviation model and the expected shortfall 
model\cite{Ciliberti}.
Pafka and Kondor 
compared the distribution of the eigenvalues of the variance-covariance matrix defined by the return rate obtained from 
the dealings market with the limit distribution acquired by assuming an independent return rate;
they also quantitatively analyzed the correlation between assets using
random matrix theory, which was developed in mathematical statistics and quantum 
chaos\cite{Pafka}. 
Shinzato and Yasuda used a belief propagation method that was developed as a decoding 
algorithm to create an algorithm which can derive the
optimal solution, with computational complexity that is proportional to the square of the number of 
investment assets\cite{Shinzato-Yasuda}. 
Wakai, Shinzato, and Shimazaki analyzed the typical behaviour when using minimal investment risk and the concentrated investment level
of Markowitz's mean-variance model for 
the cases in which the return rates of the random matrix ensemble were independently and identically 
distributed from a normal distribution, a uniform distribution, and an
exponential distribution\cite{Wakai}.

Although studies have used methods from random matrix theory and 
statistical mechanical informatics to analyze the potential risks for the portfolio optimization problem\cite{Nishimori}, 
this has been done without a mathematical proof that the 
self-averaging property of investment risk and the concentrated investment 
level can be used to effectively evaluate the optimal solution 
(self-averaging will be further discussed below)\cite{Shinzato-RIMS}. 
However, it is not obvious that these indicators are self-averaging. 
Furthermore, if they are self-averaging, the potential risk 
of an investment system can be analyzed in this way, but it is important to determine whether this is so since the results are not always in agreement with those produced using
operations research. 
Thus it is necessary to consider these problems 
systematically.

Therefore, in this paper, we provide a mathematical proof of the self-averaging property and discuss the validity of the analytical procedure that is widely used in operations research approaches to the  portfolio optimization problem. 
To do this, we reformulate the portfolio optimization problem using a probabilistic framework.  We also 
consider two scenarios, neither of which have been previously addressed, for the optimization of stochastic phenomena. 
We introduce the concept of self-averaging, and we then use it to analyze the potential risk of an investment system and determine the optimal investment strategy. 
We validate our proposed approach by comparing it with 
the results obtained by the standard operations research method 
and with those of a numerical simulation, and finally, we summarize the    
problems of using the operations research approach for a problem with this mathematical structure.

This paper is organized as follows. In the next section, we mathematically formulate the
portfolio optimization problem and discuss an easy game that optimizes stochastic phenomena; 
this presents the viewpoint 
that we will use to analyze the potential risk of an investment system. 
Section \ref{sec3} presents the concepts we will use, such as those from statistical mechanics and probabilistic 
inequalities, and summarizes the self-averaging property, which is an important feature of the optimal investment strategy. 
In Section \ref{sec4}, we present our results and compare them with those of other methods, as discussed above. 
In the final section, we present a summary and discuss areas of future work.

\section{Model setting and optimization for stochastic phenomena\label{sec2}}

\subsection{Markowitz's mean-variance portfolio selection\label{sec2.1}}
In this subsection, we present the
mean-variance model, which is one of the most commonly used models for the 
portfolio optimization problem. We begin by considering a stable investment market with 
$N$ investment outlets, where $w_k$ represents the portfolio (or investment 
ratio) of asset $k(=1,\cdots,N)$, and $x_{k\mu}'$ denotes the return rate of 
asset $k$ in scenario $\mu(=1,\cdots,p)$. 
However, for simplicity,  
we {do not} include short sales, that is, $-\infty<w_k<\infty$, and 
we {assume} that the probability distribution of the return rate is known for each asset. 
In Markowitz's mean-variance model,
given $p$ scenarios, the investment risk is defined to be the sum of squares of the difference between the gross return for a scenario $\sum_{k=1}^Nx_{k\mu}'w_k$
and its expectation $\sum_{k=1}^NE[x_{k\mu}']w_k$; 
determining an investment strategy by minimizing the risk creates a 
hedge. 
That is to say, the investment risk of a portfolio with $N$ assets ${\bf 
w}=(w_1,\cdots,w_N)^{\rm T}\in{\bf R}^N$ is defined as
\bea
{\cal H}({\bf w}|X)\eq
\f{1}{2N}
\sum_{\mu=1}^p
\left(\sum_{k=1}^Nx_{k\mu}'w_k-
\sum_{k=1}^NE[x_{k\mu}']w_k
\right)^2\nn
\eq\f{1}{2}\sum_{\mu=1}^p
\left(\f{1}{\sqrt{N}}\sum_{k=1}^Nx_{k\mu}w_k\right)^2\label{eq1},
\eea
where ${\rm T}$ denotes the transpose of a matrix or 
vector, and $E[f(x)]$ is the expectation of $f(x)$. Since we have assumed that the probability distribution of the return rate of each asset 
is known, we represent the return rate as $x_{k\mu}=x_{k\mu}'-E[x_{k\mu}']$ and the
return rate matrix as
$X=\left\{\f{x_{k\mu}}{\sqrt{N}}\right\}\in{\cal M}_{N\times p}$. 
Also, note that 
{although we introduce the coefficient $\f{1}{2N}$ in \sref{eq1} for simplicity of the discussion below,}
since $\sum_{k=1}^Nx_{k\mu}w_k$ is the summation of $N$ random variables 
$x_{k\mu}w_k$ ($w_k$ can be interpreted as the coefficient of a random variable $x_{k\mu}$),
even if we do not assume that the return rates of the assets are {independent, if ${\bf w}$ is fixed,} the correlation between the returns is small,
and the third and higher moments of the return rate are finite, 
{then} we expect that as the number of investment outlets $N$ increases, $v_\mu=\f{1}{\sqrt{N}}\sum_{k=1}^Nx_{k\mu}w_k$ 
asymptotically approaches a multidimensional Gaussian distribution {according to the central limit theorem}.

In the mean-variance model, in the absence of constraints (such as budgets), an obvious optimal portfolio is obtained by minimizing the risk function ${\cal H}({\bf w}|X)$ with 
$w_1=\cdots=w_N=0$. 
Since this is equivalent to not investing, there is no investment risk; however,
 in this paper, we {use} the budget 
 constraint 
\bea
\sum_{k=1}^Nw_k\eq N.\label{eq2}
\eea
Moreover, although in the actual management of assets, it is necessary to impose expected return restrictions in addition to budget constraints, for simplicity, we will consider only budget constraints. 
Therefore, the portfolio optimization problem is formulated as 
determining the portfolio ${\bf w}$ that minimizes ${\cal H}({\bf w}|X)$, that is, the risk function 
in \sref{eq1} with the constraint of \sref{eq2}. 
In the case of $p> N$, the optimal solution can be analytically determined: 
\bea
{\bf w}\eq\f{NJ^{-1}{\bf e}}{{\bf e}^{\rm T}J^{-1}{\bf e}}\label{eq3},
\eea
where the unit vector ${\bf e}=(1,1,\cdots,1)^{\rm T}\in{\bf R}^N$ and $J^{-1}$ is the inverse of the variance-covariance matrix $J=\left\{J_{ij}\right\}=XX^{\rm T}\in{\cal M}_{N\times N}$, where element $i,j$ of matrix $J$ is
\bea
J_{ij}\eq\f{1}{N}\sum_{\mu=1}^px_{i\mu}x_{j\mu}.\label{eq4}
\eea
If $p\le N$, then since matrix $J$ is not a regular matrix, 
the optimal solution of this portfolio optimization problem cannot be uniquely 
determined.

Using the definition of ${\cal H}({\bf w}|X)$ in \sref{eq1}, for each scenario 
 $\mu$,
we can estimate the sum of squares of the difference between the gross earnings $\sum_{k=1}^Nx_{k\mu}'w_k$ and its expectation 
$\sum_{k=1}^NE[x_{k\mu}']w_k$; this can be interpreted as the investment potential of 
 portfolio ${\bf w}$, and the  
concentrated 
investment level $q_w$ is defined as follows\cite{Ciliberti,Shinzato-Yasuda,Wakai}:
\bea
q_w\eq\f{1}{N}\sum_{k=1}^Nw_k^2\label{eq5}.
\eea
With an
equipartition investment strategy ${\bf w}=(1,1,\cdots,1)^{\rm 
T}\in{\bf R}^N$, we obtain $q_w=1$; with a 
concentrated investment strategy, for example, investing only in asset $1$,
${\bf w}=(N,0,\cdots,0)^{\rm T}\in{\bf R}^N$, so $q_w=N$ is 
obtained; if one investor invests equally in $m$ of $N$ possible outlets, 
$q_w=N/m$. Thus we have 
\bea
q_w-1\eq\f{1}{N}\sum_{k=1}^Nw_k^2-
\left(\f{1}{N}\sum_{k=1}^Nw_k\right)^2\nn
\eq\f{1}{N}\sum_{k=1}^N\left(w_k-\f{1}{N}\sum_{k'=1}^Nw_{k'}\right)^2,
\eea
and as portfolio ${\bf w}$ approaches equipartition, $q_w$ decreases to $1$, and as it approaches a 
concentrated investment strategy, $q_w$ increases.

We note that although {
$\sum_{k=1}^Nw_k=1$ is widely used as a budget constraint in operations research,} 
we do not use one in this paper.
Since the optimal solution to the portfolio optimization problem with a
budget constraint that is widely used in operations research is ${\bf 
w}^1=(w_1^1,w_2^1,\cdots,w_N^1)^{\rm T}\in{\bf R}^N$, and the optimal 
solution defined in {\sref{eq2} is} ${\bf w}^N=(w_1^N,w_2^N,\cdots,w_N^N)^{\rm 
T}\in{\bf R}^N$, the relation 
$w_i^1/w_j^1=w_i^N/w_j^N$ is proved, that is, in the optimal portfolios of each method, the 
investment ratios are in agreement. Furthermore, the
concentrated investment level $q_w$ can be interpreted as 
an indicator of diversification when using the 
budget constraint of \sref{eq2}.

\subsection{Optimization for Stochastic Phenomena\label{sec2.2}}
In this subsection, we consider this {optimization problem} from a different viewpoint.  We analyze the behaviour of minimal investment risk $\ve$ and the
concentrated investment level $q_w$ of the mean-variance model, and we discuss the optimization of  stochastic 
phenomena which have not been addressed in the operations research approach to this problem. Let us consider the following variant of the well-known game of 
rock-paper-scissors. 
Rule 1: Two subjects, Alice and Bob, play rock-paper-scissors 
$300$ times. 
Rule 2: Alice can freely choose to display rock, paper, or scissors. 
On the other hand, Bob's choice is randomly assigned by the toss of a fair dice: {rock} when the dice shows $1$ or $2$,  
{paper} for $3$ or $4$, and {scissors} for $5$ or $6$. Moreover, Alice knows that Bob's choice is randomly and independently 
determined by the dice. Rule $3$: The 
winner adds a point, the loser subtracts a point, and if they tie, there is no change to the score. 
We now consider whether 
it is expected that Alice will win overall.

\paragraph{(a) Two subjects simultaneously hold out their hands to indicate rock, paper, {or scissors}}
First, we consider the ordinary case.
Since Alice does not know Bob's choice, she assumes that each possibility has equal probability. Thus, if Alice also chooses according to a roll of a dice,
the expected total acquired score would be $0$. Similarly, if 
	   Alice chooses only {rock}, 
the expected score would be $0$.
We will use the following notation: {
$r_A$ (resp. $r_B$) is the probability Alice (resp. Bob) chooses rock,
$p_A$ (resp. $p_B$) is the probability Alice (resp. Bob) chooses paper, and  	  
$s_A$ (resp. $s_B$) is the probability Alice (resp. Bob) chooses scissors;} note that for each player, the expectation of the total acquired score is $0$. 
That is, if Alice does not have prior knowledge of Bob's choice, neither of them can win (for a sufficient number of trials), and the expectation is that they tie.
However, if the probabilities of Bob's choice 
	   are not equal {(e.g., $(r_B,p_B,s_B)=(2/3,1/6,1/6)$), then Alice should choose $(r_A,p_A,s_A)=(0,1,0)$.} In this case, the expected total acquired score for Alice 
	   is $150$. 
Generally speaking, even if Alice does not have prior knowledge of Bob's choice, 
if she knows the probabilities of his choices, she can choose in such a 
way that maximizes her {expected} score. 
\paragraph{(b) Alice has prior knowledge of {Bob's choice}}
We now consider the case where Alice makes her choice after learning what Bob will display. Her goal is to maximize the expectation of her total score.
With added constraints, her expected total 
	  score will be larger than $0$; without constraints, it will be $300$. 
	   
\paragraph{(c) There is a constraint on the number of times that rock, paper, and scissors can {each be chosen}}
We now consider the case where Alice's choices are constrained; 
	   for instance, they must each be chosen an equal number of times (i.e., $100$ times). With this constraint, the expected total acquired score is $0$ for case (a) (Alice does not have prior knowledge of Bob's choice), but for case (b) (Alice has prior knowledge of Bob's choice), it is $500/3$. 
That is, if she has prior knowledge, she can take protective action. 
\paragraph{(d) Five sets of $300$ {sessions}}
Finally, we consider the case that two subjects play five sets of $300$ 
	   games. 
If Alice has no prior knowledge of Bob's choice, her expected total acquired score is 
	   again $0$. If she has prior knowledge and there are no constraints, 
her expected score is $1500$. 
If there is a constraint such that Alice must make the same choice each time, her expected score 
	   is $0$ for case (a) (Alice does not have prior knowledge of Bob's choice), but 
	  $5000/9$ for case (b) (Alice has prior knowledge of Bob's choice).
In case (c) (Alice's choices are constrained), 
it is easy to see that the expectation of Alice's total score for case (a) is not larger than 
it is for case (b).

In conclusion, for both case
(c) (Alice's choices are constrained) and case (d) {(five sets of 300 games and Alice makes the same choice each time)},
if Alice has prior knowledge of Bob's choice, her score will be higher than if she has no such knowledge.
That is, if Alice has prior knowledge, she can produce a better strategy.

We would like to make one more point, which will be further discussed below. 
Cases (a) and (b) (respectively, Alice does not or does have prior knowledge of Bob's choice) are similar to the 
discussion of annealed and quenched disorder systems 
in statistical mechanics\cite{Ma,Nishimori}. 
In an annealed disorder system, 
the indicator $f({\bf w}|X)$ is first averaged using a random $X$ 
in the disordered system, 
and then the averaged indicator $E[f({\bf w}|X)]$ is optimized in order to 
assess the behaviour of the system. 
In the rock-paper-scissors example, 
the indicator $f({\bf w}|X)$ corresponds to the total acquired score 
of Alice, ${\bf w}$ corresponds to Alice's choices (or 
strategy), the random $X$ corresponds to Bob's choices, and 
case (a) corresponds to the annealed disorder 
system. 
On the other hand, 
in a quenched disorder system, 
$f({\bf w}|X)$ is {first} optimized subject to {a restriction 
and a random $X$ which is included in the 
disordered system, and then the optimized indicator is 
averaged} using the random $X$ in order to 
assess the behaviour of this system; this corresponds to case (b) of 
the rock-paper-scissors example. 
More generally, 
when optimizing an indicator $f({\bf w}|X)$ for a stochastic phenomena,
it matters in which order the averaging and optimizing occur.
When maximizing, it is necessary to precisely estimate two kinds of indicators, $f_a=
\mathop{\max}_{\bf w}E[f({\bf w}|X)]$ and $
f_q=E\left[\mathop{\max}_{\bf w}f({\bf w}|X)\right]
$.
Since for any ${\bf w}$, $\mathop{\max}_{\bf w}f({\bf w}|X)\ge f({\bf w}|X)$ holds
for any random $X$, to find the relative magnitude of $f_a$ and $f_q$,
one averages both sides and then maximizes the 
right-hand side. The left-hand side does not need to be maximized because a definite value is obtained for $f_q\ge f_a$. When minimizing, we have
$f_a=\mathop{\min}_{\bf w}E[f({\bf w}|X)]$
 and $f_q= E\left[\mathop{\min}_{\bf w}f({\bf w}|X)\right]$, and so in way similar to the above, we obtain $f_a\ge 
f_q$, that is, 
\bea
\mathop{\min}_{\bf w}E[f({\bf w}|X)]
\ge E\left[\mathop{\min}_{\bf w}f({\bf w}|X)\right].\label{eq7}
\eea
\subsection{Operations research approach for portfolio optimization\label{sec2.3}}
From the above argument, 
we see that optimization of stochastic phenomena is handled differently for an annealed disorder system 
than it is for a quenched disorder system. 
Using this core concept, let us reconsider the portfolio optimization problem. In the standard analytical approach of operations research 
to the portfolio optimization problem, 
one first averages the risk function ${\cal H}({\bf w}|X)$ with the return 
rate on the assets and then minimizes the expectation of the risk function $E[{\cal H}({\bf w}|X)]$ with a budget 
constant. Here, for simplicity, 
we presume that return rate is independently and identically 
distributed with a standard normal distribution. Thus, the expectation 
of the correlation between asset $i$ and asset $j$ is 
\bea
E[J_{ij}]\eq
\left\{
\begin{array}{ll}
\f{p}{N}&i=j\\
0&i\ne j.
\end{array}
\right.
\eea
Using this, the expected investment risk function $E[{\cal 
H}({\bf w}|X)]$ with a return rate on the assets is 
\bea
E[{\cal H}({\bf w}|X)]\eq\f{\a}{2}\sum_{k=1}^Nw_k^2,\label{eq9}
\eea
where {the ratio $\a=p/N$ is used}. In addition, from the
symmetry of this model, the optimal investment strategy of \sref{eq9}, using the budget 
constraint of \sref{eq2}, describes an equipartition investment strategy. The 
minimum expected investment risk per asset $\ve^{\rm OR}
=
\f{1}{N}
\mathop{\min}_{\bf w}E[{\cal H}({\bf w}|X)]
$ is evaluated 
as follows:
\bea
\ve^{\rm OR}\eq
\f{\a}{2}.
\eea
The concentrated investment level $q_w^{\rm OR}$ is 
\bea
q_w^{\rm OR}\eq1.
\eea
This analytical approach, 
which is widely used in operations research, 
does not provide insight into the optimal investment 
strategy in an actual market; the reason is not just that the model was simplified by assuming the return rate is 
independently and identically distributed with a standard normal distribution. 
Since this analytical approach is equivalent to case (a) in the
rock-paper-scissors game with two subjects and the 
annealed 
disorder system, it is not clear that 
this approach could be used to minimize the 
investment risk ${\cal H}({\bf w}|X)$ 
with respect to a realistic individual return rate matrix $X$, 
that is, ${\bf w}=\arg\mathop{\min}_{{\bf w}}{\cal H}({\bf w}|X)$. 
In particular, the equality $
\arg\mathop{\min}_{\bf w}{\cal H}({\bf w}|X)=
\arg\mathop{\min}_{\bf w}E[{\cal H}({\bf w}|X)]
$
is not always satisfied. 
As we discussed with the rock-paper-scissors game, 
if we average the investment risk with the return rate, we can avoid the complication of optimizing 
individual return rates; 
on the other hand, this approach does not evaluate the optimal strategy 
based on individual return rates. Even though 
it is not guaranteed mathematically that the 
solution to the minimal expected investment risk 
optimizes the investment risk for each set of return 
rates ${\cal H}({\bf w}|X)$, 
this approach is widely used in operations research and might provide a misleading investment strategy.

On the other hand, let us consider case
(b) in the rock-paper-scissors game with two subjects and the quenched disorder system. 
In a stable investment market, even if at $\mu=0$
one had prior information about the probability distribution of the return rate of each asset
during the next period ($\mu=1$ to $\mu=p$), 
since it is not possible to know the actual return rate, 
it is difficult to select the optimal investment strategy. 
However, if we have prior information about the return rates, 
as discussed in the rock-paper-scissors example, we can minimize the 
investment risk and obtain an optimal investment strategy. 
In particular, if $p/N\le1$, since it is well known that the
optimal solution is a linear sum of the eigenvectors of the minimal 
eigenvalue of the variance-covariance matrix $J$, {the investment 
risk per asset} is 
$0$, 
since the minimal eigenvalue of the matrix $J$ is $0$ since $J$ is nonsingular. We next consider the
concentrated investment level $q_w$. Let ${\cal V}$ be {variance of} the sample variance 
$\f{1}{p}\sum_{\mu=1}^px_{k\mu}^2$; it is evaluated as follows: 
\bea
{\cal V}\eq
E
\left[
\left(\f{1}{p}\sum_{\mu=1}^px_{k\mu}^2-
\f{1}{p}\sum_{\mu=1}^pE[x_{k\mu}^2]
\right)^2\right]\nn
\eq\f{2}{N\a}.
\eea
When $\a=p/N$ is small, ${\cal V}$ is large, and  
one should invest heavily in blue-chip assets for which the return rates have smaller sample variances than those of the other $N$ investment outlets; in this way, the risk is decreased, and 
the optimal 
investment strategy is asymptotically close to a concentrated investment 
strategy, namely $q_w\gg1$.

When $p/N>1$,
using the optimal solution for \sref{eq3}, the two 
indicators can be analytically assessed. That is, the minimal investment risk 
per asset $\ve(X)$ and the concentrated investment level 
$q_w(X)$ can be written as follows:
\bea
\label{eq13}
\ve(X)\eq
\f{1}{N}{\cal H}\left(\left.\f{NJ^{-1}{\bf e}}{{\bf e}^{\rm T}J^{-1}{\bf e}}
\right|X\right)\nn
\eq\f{N}{2\left({\rm e}^{\rm T}J^{-1}{\bf e}\right)},\\
q_w(X)\eq\f{1}{N}\left(
\f{NJ^{-1}{\bf e}}{{\bf e}^{\rm T}J^{-1}{\bf e}}
\right)^{\rm T}
\left(\f{NJ^{-1}{\bf e}}{{\bf e}^{\rm T}J^{-1}{\bf e}}\right)\nn
\eq\f{N{\bf e}^{\rm T}J^{-2}{\bf e}}
{\left(
{\bf e}^{\rm T}J^{-1}
{\bf e}
\right)^2
}\label{eq14},
\eea
where we use the explicit return rate matrix $X=\left\{\f{x_{k\mu}}{\sqrt{N}}\right\}\in{\cal 
M}_{N\times p}$ as the argument since these indicators depend 
on the return rate matrix. The
variance-covariance matrix $J=XX^{\rm T}\in{\cal M}_{N\times N}$ has already 
been defined. 
In actual investments, 
assuming fair dealing, 
since we do not have prior knowledge of the actual return rate, we cannot precisely determine the two indicators. 
However, we can evaluate the previous risk in an
investment system 
and thus support the strategy of an investor.
In order to provide useful insight, we need to precisely analyze $\ve(X)$ and $q_w(X)$. 
For the reasons noted here, we assume that during the
initial period, we have prior knowledge of the return rate; although this assumption is impossible, we note that we will show below that this assumption is not 
required to evaluate the potential of an investment system. 
Although we need to assess the optimal solution or the inverse of the
variance-covariance matrix in order to assess the potential risk 
of an investment market, it is difficult to do this
since the computational complexity of finding the inverse matrix is proportional to the 
cube of the matrix size $N$, which is the number of 
investment outlets. 
In addition, we need to find each inverse matrix for each return rate 
in order to evaluate the minimal investment risk of each set; 
however, if the minimal investment risk randomly fluctuates with the
return rates, we would need to average the minimal investment risk with the
return rate set $X$. 
We now note that we can use the  
self-averaging property to simplify evaluation of the potential investment risk.

\section{Preliminaries\label{sec3}}
We first prepare some mathematical tools to enable discussion of the
self-averaging property of the minimal investment risk. 
\subsection{Statistical mechanics\label{sec3.1}}
First, using the Boltzmann distribution of the inverse temperature $\b(>0)$, which is 
widely used in statistical mechanics, 
the posterior probability of portfolio ${\bf w}$ given return rate 
matrix $X$, $P({\bf 
w}|X)$, is defined as follows:
\bea
P({\bf w}|X)\eq\f{P_0({\bf w})e^{-\b{\cal H}({\bf w}|X)}}
{Z(\b,X)}\label{eq16},
\eea
where the prior probability $P_0({\bf w})$ is $1$ if portfolio ${\bf 
w}$ satisfies \sref{eq2}, and it is $0$ otherwise; 
$e^{-\b{\cal H}({\bf w}|X)}$ is the likelihood function; and 
$Z(\b,X)$, the 
partition function, is a normalized 
constant and is defined as follows:
\bea
Z(\b,X)\eq
\area d{\bf w}P_0({\bf w})e^{-\b{\cal H}({\bf w}|X)}\label{eq17}.
\eea
From this, it is found that the posterior probability $P({\bf w}|X)$ satisfies the 
property of a probability measure, that is, 
$P({\bf w}|X)\ge0$ and $\area d{\bf w}P({\bf w}|X)=1$.
Furthermore, 
it is well known that ${\bf w}^*=\arg\mathop{\max}_{\bf w}P({\bf w}|X)$, which is obtained 
using the maximum a posteriori estimation, 
is consistent with 
the portfolio obtained by minimizing the investment risk function ${\cal H}({\bf w}|X)$.
By taking the limit of the inverse temperature $\b$, we obtain 
\bea
\lim_{\b\to\infty}P({\bf w}|X)
\eq\prod_{i=1}^N\d(w_i-w_i^*),\label{eq17-1}
\eea
where $w_i^*$ is the optimal investment ratio for asset 
$i$. Thus, we can average the portfolio ${\bf w}$ and the investment risk ${\cal H}({\bf w}|X)$
using the a posteriori 
probability $P({\bf w}|X)$ and allow the inverse 
temperature $\b$ to become sufficiently large:
\bea
{\bf w}^*\eq\lim_{\b\to\infty}\area d{\bf w}P({\bf w}|X){\bf w}\\
{\cal H}({\bf w}^*|X)\eq\lim_{\b\to\infty}\area d{\bf w}P({\bf 
w}|X){\cal H}({\bf w}|X),
\eea
where $\d(u)$ is the Dirac delta function and this holds
for any function $f(x)$ such that
$f(x)=\area dyf(y)\d(y-x)$ (see \ref{app2}). 
From this reformulation and
using the posterior probability defined in \sref{eq16},
the portfolio optimization problem can be solved using the framework of 
{probabilistic reasoning}.

\subsection{Chernoff inequality}
Next, we introduce {one of the probability inequalities}, the Chernoff inequality, as follows.
For a random variable $Y$ with known probability measure and 
a constant number $\eta$, the 
probability {that} $\eta\le Y$  
satisfies the following inequality for any $u>0$ is\cite{Chernoff,Gallager}: 
\bea
Pr[\eta\le Y]&\le&e^{-u\eta}E[e^{uY}]\label{eq21}.
\eea
This can be easily proved; for example, 
consider the step function $\Theta(W)$, which is $1$ if $W\ge0$ and $0$ otherwise. 
First, for $u>0$, we obtain
$\Theta(W)\le e^{uW}$. From this, we derive
$Pr[\eta\le Y]=E[\Theta(Y-\eta)]\le e^{-u\eta}E[e^{uY}]$. In 
addition, for $Pr[\eta\ge Y]$, we obtain $Pr[\eta\ge Y]\le 
e^{-u\eta}E[e^{uY}]$ for $u<0$.

From \sref{eq21},
we could derive a tighter upper bound.
Since the right-hand side in \sref{eq21}
is guaranteed for an arbitrary $u>0$, 
there necessarily exists a minimum value for the right-hand side for any $u>0$, and we obtain
\bea
Pr[\eta\le Y]&\le&\mathop{\min}_{u>0}\left\{e^{-u\eta}E[e^{uY}]\label{eq22}\right\}\nn
\eq e^{-R(\eta)}.
\eea
Here $R(\eta)$ is the rate function and is defined as
\bea
R(\eta)\eq\mathop{\max}_{u>0}\left\{u\eta-\log E[e^{uY}]\right\}.
\eea
The cumulative generating function 
$\phi(u)=\log E[e^{uY}]$ is a convex function of $u$, and the 
rate function $R(\eta)$, defined by the Legendre 
transformation of a convex function, is also a convex function.
It is also known that $R(\eta)$ is nonnegative, $R(\eta)=0$ if $\eta\le E[Y]$, and 
$R(\eta)>0$ if $\eta>E[Y]$. These properties of the rate function are proved in 
 \ref{app1}.
\subsection{Self-averaging property supported by large deviation theory}
When the portfolio 
${\bf w}$
depends on the posterior probability 
$P({\bf w}|X)$ defined in 
\sref{eq16},
the probability that the investment risk per asset $\f{1}{N}{\cal H}({\bf w}|X)$
is less than or equal to a constant number $\tilde{\ve}$ satisfies the Chernoff inequality; that is, $Pr\left[\f{1}{N}{\cal H}({\bf 
w}|X)\le\tilde{\ve}\right]=E\left[\Theta\left(N\tilde{\ve}-{\cal H}({\bf 
w}|X)\right)\right]
=\area d{\bf w}P({\bf w}|X)\Theta\left(N\tilde{\ve}-{\cal H}({\bf 
w}|X)\right)
$ satisfies
\bea
\label{eq24}
Pr\left[\f{1}{N}{\cal H}({\bf w}|X)\le\tilde{\ve}\right]&\le&
E\left[e^{N\tilde{\b}\tilde{\ve}-\tilde{\b}{\cal H}({\bf w}|X)}\right]\nn
\eq e^{N\tilde{\b}\tilde{\ve}}
\f{Z(\b+\tilde{\b},X)}{Z(\b,X)}.
\eea
Here, $\tilde{\b}$ is a positive number and $Z(\b+\tilde{\b},X)$ is defined by \sref{eq17} (with $\b$ replaced by $\b+\tilde{\b}$). Thus, the
probability inequality of the tighter upper bound  of \sref{eq24} is derived 
using the following rate function:
\bea
R_+({\b},\tilde{\ve},X)
\eq\mathop{\max}_{\tilde{\b}>0}
\left\{
-\tilde{\b}\tilde{\ve}-\f{1}{N}\log Z(\b+\tilde{\b},X)
+\f{1}{N}\log Z(\b,X)
\right\}\label{eq25},
\eea
and we have  
$Pr\left[\f{1}{N}{\cal H}({\bf w}|X)\le\tilde{\ve}\right]\le e^{-NR_+(\b,\tilde{\ve},X)}$\cite{Touchette}. 
In a similar way, {for} the probability of 
$\f{1}{N}{\cal H}
({\bf w}|X)\ge\tilde{\ve}$, $Pr\left[\f{1}{N}{\cal H}
({\bf w}|X)\ge\tilde{\ve}
\right]$, {thus}
$Pr\left[\f{1}{N}{\cal H}
({\bf w}|X)\ge\tilde{\ve}
\right]\le e^{-NR_-(\b,\tilde{\ve},X)}$ is {also} obtained, where we use the rate function 
\bea
R_-(\b,\tilde{\ve},X)\eq\mathop{\max}_{\tilde{\b}<0}
\left\{
-\tilde{\b}\tilde{\ve}-\f{1}{N}\log Z(\b+\tilde{\b},X)
+\f{1}{N}\log Z(\b,X)
\right\}.\label{eq26}
\eea
In order to analyze the rate functions in \sref{eq25} and 
\sref{eq26}, it is {also} necessary to assess 
$\f{1}{N}\log Z(\b+\tilde{\b},X)$ and 
$\f{1}{N}\log Z(\b,X)$, which depend on the return rate matrix $X$. 
Based on the definition in \sref{eq17}, 
assessing these partition functions analytically
 is more difficult than assessing the optimal solution analytically.
In order to resolve this difficulty, 
we consider the cumulative distribution of $\f{1}{N}\log Z(\b,X)$ or the Helmholtz free energy 
$f(\b,X)$, as defined in the following equation: 
\bea
f(\b,X)\eq-\f{1}{N\b}\log Z(\b,X)\label{eq27}.
\eea
The Helmholtz free energy $f(\b,X)$ fluctuates randomly 
with the probability of the return rate matrix $X$. Thus, it is necessary to evaluate the
Chernoff inequality for the Helmholtz free 
energy and its rate function: 
\bea
Pr\left[f(\b,X)\le\tilde{f}\right]&\le& e^{Nn\b\tilde{f}}E[e^{-Nn\b f(\b,X)}]\nn
\eq e^{Nn\b\tilde{f}}E[Z^n(\b,X)],
\eea
where $n>0$ has already been defined. In a similar way, we have
\bea
Pr\left[f(\b,X)\ge\tilde{f}\right]&\le&e^{Nn\b\tilde{f}}E[Z^n(\b,X)],
\eea
where $n<0$. In conclusion, we obtain the
two probability inequalities, 
$Pr\left[f(\b,X)\le\tilde{f}\right]\le e^{-NR_+(\b,\tilde{f})}$ and $Pr\left[f(\b,X)\ge\tilde{f}\right]\le e^{-NR_-(\b,\tilde{f})},$
where 
\bea
\label{eq30}R_+(\b,\tilde{f})\eq\mathop{\max}_{n>0}\left\{-n\b\tilde{f}-\f{1}{N}\log 
E[Z^n(\b,X)]\right\},\\
R_-(\b,\tilde{f})\eq\mathop{\max}_{n<0}\left\{-n\b\tilde{f}-\f{1}{N}\log 
E[Z^n(\b,X)]\right\}\label{eq31}.
\eea
In both inequalities, it is necessary to 
analyze $\f{1}{N}\log E[Z^n(\b,X)]$. 
\section{Replica analysis and numerical simulation\label{sec4}}
\subsection{Similarity to the Hopfield model}
In this subsection, 
in order to determine whether we can use replica analysis to evaluate
$E[Z^n(\b,X)]$, 
let us consider briefly the problem of recalling a pattern stored in a neural network 
constructed of $N$ neurons; the mathematical structure of this problem is similar to the
portfolio optimization problem\cite{Amit87,Nishimori}.
Let $S_k$ be the state of neuron $k$; then $S_k=1$ if neuron $k$ has been fired 
and $S_k=-1$ otherwise. Additionally, 
$x_{k\mu},(k=1,\cdots,N,\mu=1,\cdots,p)$ is the memory of neuron $k$ 
for pattern $\mu$ included in $p$ stored patterns, and it is randomly 
assigned $\pm1$ with equal probability.

Then, for $p$ patterns, the Hebb rule is defined as follows:
\bea
J_{ij}\eq\f{1}{N}\sum_{\mu=1}^px_{i\mu}x_{j\mu}
\eea
where {$J_{ij}$} is the correlation between neuron $i$ and neuron $j$.
Thus, it is well known that the
neuron state ${\bf S}$ that minimizes the Hamiltonian ${\cal H}({\bf S}|X)$ in \sref{eq33}
is consistent with each stored pattern:
\bea
\label{eq33}{\cal H}({\bf S}|X)\eq-\f{1}{2}{\bf S}^{\rm T}J{\bf S}.
\eea
If neuron state ${\bf S}$ is consistent with pattern $1$, 
that is, $S_k=x_{k1}$, the
Hamiltonian ${\cal H}({\bf S}|X)$ can be written as 
\bea
{\cal H}({\bf S}|X)\eq-\f{1}{2N}\left(\sum_{k=1}^Nx_{k1}S_k\right)^2
-\f{1}{2N}\sum_{\mu=2}^p\left(\sum_{k=1}^Nx_{k\mu}S_k\right)^2\nn
\eq-\f{N}{2},
\eea
 where the
overlap between pattern $\mu$ and pattern $\nu$ in limited by the 
number of neurons $N$ and satisfies
\bea
\f{1}{N}\sum_{k=1}^Nx_{k\mu}x_{k\nu}\eq
\left\{\begin{array}{ll}
1&\mu=\nu\\
0&\mu\ne\nu
\end{array}
\right..
\eea
Intuitively, each pattern is orthogonal with each of the others, since the stored 
patterns are independent and are randomly assigned. 
This problem of recalling patterns stored in a neural 
network and of accounting for the number of identifiable patterns 
is called the associative memory problem, and the model defined in 
\sref{eq33} is called the Hopfield model.

In the analysis of the Hopfield model, the
upper limit of the number of identifiable patterns 
is estimated using $E[Z^n(\b,X)]$, which evaluates the 
learning potential of the neural network. We also 
note the mathematical similarity between this model and the mean-variance model, which indicates that 
we could adapt the analytical approach used for the Hopfield model; that is, we could use
replica analysis for the portfolio optimization problem 
and to assess the potential of the investment system.

\subsection{Main results obtained in replica analysis\label{sec4.2}}
For the detailed calculations of replica analysis, please see \ref{app2}.  We will limit the number of investment outlets $N$ such that $\a=p/N\sim O(1)$. For $n\in{\bf N}$, we have
\bea
\Phi(n)\eq\lim_{N\to\infty}\f{1}{N}\log E[Z^n(\b,X)]\nn
\eq\mathop{\rm Extr}_{{\bf k},Q_w,\tilde{Q}_w}
\left\{
-\f{\a}{2}\log\det\left|I+\b Q_w\right|
+\f{1}{2}{\rm Tr}Q_w\tilde{Q}_w
-\f{1}{2}\log\det\left|\tilde{Q}_w\right|
\right.\nn
&&\left.
-{\bf e}^{\rm T}{\bf k}+\f{1}{2}{\bf k}^{\rm 
T}\tilde{Q}_w^{-1}{\bf k},
\right\}\label{eq36}
\qquad\qquad
\eea
where ${\bf k}=(k_1,\cdots,k_n)^{\rm T}\in{\bf R}^n,Q_w=\left\{q_{wab}\right\}\in{\cal M}_{n\times 
n}, \tilde{Q}_w=\left\{\tilde{q}_{wab}\right\}\in{\cal M}_{n\times 
n}$, $k_a,q_{wab}$, and $\tilde{q}_{wab}$ are order parameters, 
${\bf 
e}=(1,\cdots,1)^{\rm T}\in{\bf R}^n$ is a constant vector, $I\in{\cal 
M}_{n\times n}$ is the identity matrix, and $\mathop{\rm Extr}_{A}f(A)$ 
are the extrema of $f(A)$ with respect to $A$. From this, the
extrema of {${\bf k},Q_w$, and $\tilde{Q}_w$} are assessed as follows:
\bea
{\bf k}\eq\tilde{Q}_w{\bf e},\\
\tilde{Q}_w\eq\b(\a-1)I-\f{\b^2(\a-1)}{1+n\b}D,\\
Q_w\eq\f{1}{\b(\a-1)}I+\f{\a}{\a-1}D,
\eea
where $D={\bf e}{\bf 
e}^{\rm T}\in{\cal 
M}_{n\times n}$ is a square matrix, all of whose components are $1$. 
Based on these results, 
we do not need to assume replica symmetry ansatz with respect to the order 
parameters of this model ($k_a,q_{wab},\tilde{q}_{wab}$). Thus, 
substituting these result into \sref{eq36}, we have
\bea
\Phi(n)\eq
-\f{n\a}{2}\log\f{\a}{\a-1}-\f{\a-1}{2}\log\left(1+n\b\right)+\f{n}{2}
-\f{n}{2}\log\b(\a-1)\label{eq40}.
\eea
Here we should note that in
\ref{app2}, we require that the replica number $n$ in \sref{eq36} is a natural 
number, that is, since $E[Z^n(\b,X)]$ at replica number $n\in{\bf N}$ can be estimated 
comparatively easily, the replica number $n$ in \sref{eq40} should also be a natural 
number. However, in the optimization in \sref{eq30} and \sref{eq31}, 
{we} need to have $n\in{\bf R}$ to adequately discuss the solution. 
Thus, we assume here that the replica number $n$ in \sref{eq40} is a real 
number and use this to discuss our approach in detail. 
In Subsection \ref{sec4.4}, we will compare this result with the 
result to justify that this is applicable.

The two rate functions are calculated as follows:
\bea
R_+(\b,\tilde{f})\eq
\left\{
\begin{array}{ll}
0&\f{\a-1}{2}-\f{\Lambda(\b)}{2\b}\le\tilde{f}\\
\f{\a-1}{2}\left(s-1
-\log s
\right)&\f{\a-1}{2}-\f{\Lambda(\b)}{2\b}>\tilde{f}
\end{array}
\right.\\
R_-(\b,\tilde{f})\eq
\left\{
\begin{array}{ll}
\f{\a-1}{2}\left(s-1-\log s\right)
&\f{\a-1}{2}-\f{\Lambda(\b)}{2\b}<\tilde{f}\\
0&\f{\a-1}{2}-\f{\Lambda(\b)}{2\b}\ge\tilde{f},
\end{array}
\right.
\eea
where 
\bea
\Lambda(\b)\eq1-\a\log\f{\a}{\a-1}-\log\b(\a-1)\\
s\eq\f{\tilde{f}+\f{\Lambda(\b)}{2\b}}
{\f{\a-1}{2}}.
\eea
This result satisfies the 
properties of a rate function, as shown in \ref{app1}. Moreover, 
using Gibbs inequality, $s-1-\log s\ge0$,
if $\a>1$, in the limit as the number of investment outlets $N$ becomes 
sufficiently large, we obtain
\bea
\label{eq44}
Pr\left[f(\b,X)\le\tilde{f}\right]\eq
\left\{
\begin{array}{ll}
1&\f{\a-1}{2}-\f{\Lambda(\b)}{2\b}\le\tilde{f}\\
0&\f{\a-1}{2}-\f{\Lambda(\b)}{2\b}>\tilde{f}
\end{array}
\right.\\
Pr\left[f(\b,X)\ge\tilde{f}\right]\eq
\left\{
\begin{array}{ll}
0&\f{\a-1}{2}-\f{\Lambda(\b)}{2\b}<\tilde{f}\\
1&\f{\a-1}{2}-\f{\Lambda(\b)}{2\b}\ge\tilde{f}
\end{array} .
\right.\label{eq45}
\eea
From \sref{eq44} and \sref{eq45}, since 
$f(\b,X)$ is localized around the constant 
$\f{\a-1}{2}-\f{\Lambda(\b)}{2\b}$, 
\bea
f(\b,X)\eq
\f{\a-1}{2}-\f{\Lambda(\b)}{2\b}
\eea
is verified for a realistic set of return rates. 
Namely, the Helmholtz free energy per asset $f(\b,X)$, which is a function of the
random variable $X$, becomes a definite value 
in the limit of sufficiently large $N$. 
Thus, we have
\bea
f(\b,X)\eq E[f(\b,X)].\label{eq48}
\eea
In addition, because of localizing around the definite value, 
$f^m(\b,X)=E[f^m(\b,X)]$ is also satisfied. 
This 
property in which a statistic or function of a random variable 
localizes around a definite value (or its average) is called a self-averaging property. 
By substituting \sref{eq48} into \sref{eq27}, we obtain
\bea
\f{1}{N}\log Z(\b,X)\eq
\f{\Lambda(\b)}{2}
-\f{\b(\a-1)}{2} ,
\eea
and the two rate functions 
\bea
R_+(\b,\tilde{\ve},X)\eq
\left\{
\begin{array}{ll}
+\infty&\tilde{\ve}\le\f{\a-1}{2}\\
\f{1}{2}({s'-1-\log s'})&\f{\a-1}{2}<\tilde{\ve}<\f{\a-1}{2}+\f{1}{2\b}\\
0&\f{\a-1}{2}+\f{1}{2\b}\le\tilde{\ve}
\end{array}
\right.\\
R_-(\b,\tilde{\ve},X)\eq
\left\{
\begin{array}{ll}
\f{1}{2}({s'-1-\log s'})&\f{\a-1}{2}+\f{1}{2\b}<\tilde{\ve}\\
0&\tilde{\ve}\le\f{\a-1}{2}+\f{1}{2\b}
\end{array} ,
\right.
\eea
where {$s'=2\b\left(\tilde{\ve}-\f{\a-1}{2}\right)$.} 
Thus, for a sufficiently large $N$, since the investment 
risk per asset $\f{1}{N}{\cal H}({\bf w}|X)$ is also localized around $\f{\a-1}{2}+\f{1}{2\b}$,
the investment risk is self-averaging. 
Moreover, for a sufficiently large $\b$, 
from \sref{eq17-1} 
we can derive the minimal investment risk, as follows: 
\bea
\ve(X)\eq
\f{\a-1}{2}\label{eq52}.
\eea

Furthermore, since $\ve(X)$ is also derived analytically from an identical equation, $\ve(X)=-\lim_{\b\to\infty}\f{1}{N}\pp{}{\b}\log 
Z(\b,X)$, we have validated our method in another way (see 
\ref{app2}). In addition, from the self-averaging property 
of the investment risk, 
since we can ignore the dependency of the investment risk $\ve(X)$ on the return rate matrix $X$, we will replace $\ve$ with $\ve(X)$. In 
a similar way, $q_w(X)$ is also self-averaging,  and so
then 
\bea
q_w\eq\f{\a}{\a-1}
\eea
is obtained, where $q_w(X)$ has been replaced by $q_w$.

We also note that since the minimal investment risk per asset and the concentrated investment 
level are both self-averaging (since their dependency 
on the return rate matrix $X$ is ignored), we can estimate the potential of 
this investment system. In a stable investment market, this 
implies that the
minimal investment risk with respect to a realistic return rate averaged over an investment period and 
the minimal investment risk defined by the 
return rate are in agreement since the
minimal investment risk is self-averaging.
Because of this, 
we do not need {the assumption of the quenched disorder system that 
during the initial period, we have prior knowledge of the return rates 
that is, we only need to know a priori the previous return rates.
This is another advantage of the self-averaging property.}

\subsection{Comparison with results obtained by the operations research approach}
In this subsection, we compare the two indicators that were derived in 
subsection \ref{sec2.3} using the analytical approach of operations research. 
For example, we consider  $\ve^{\rm 
OR}=\f{\a}{2}$ and $q_w^{\rm OR}=1$ with the two feature indicators derived in subsection \ref{sec4.2}
and using the self-averaging property, that is, $\ve=\f{\a-1}{2}$ if $\a>1$ 
and $\ve=0$ otherwise, and $q_w=\f{\a}{\a-1}$ if $\a>1$ and $q_w\gg1$ otherwise.
Thus, for any $\a$, we have
\bea
\label{eq52-1}\ve^{\rm OR}&\ge&\ve\\
q_w^{\rm OR}&\le&q_w.
\eea
First, the minimal expected investment risk $\ve^{\rm OR}$ 
is not smaller than the expected minimal investment risk $\ve$, {that is, \sref{eq52-1}} is consistent with the relationship in \sref{eq7}. 
Next, from both of the concentrated 
investment levels and using the 
analytical procedure of operations research, we find that the risks for each
investment outlet are averaged and negated by the returns matrix. We thus conclude that the optimal strategy is 
equipartition investing. 
On the other hand, 
when using our proposed 
method, since 
{it is possible to find the optimal solution for each investment 
outlet}, the best return rate is found for an
investment outlet that has little variation, 
especially if $\a$ is small, 
and this implies that the optimal strategy is concentrated investing.

Furthermore, 
we provide another intuitive interpretation  
using another mathematical argument.
By the definition of $q_w$ and the $N$ eigenvalues of matrix $J=XX^{\rm 
T}\in{\cal M}_{N\times N}$, 
$\l_k,(k=1,\cdots,N,\l_1\le\l_2\le\cdots\le\l_N)$, then 
$q_w=E[\l^{-2}]/(E[\l^{-1}])^2$ and 
where $E[\l^{-s}]=\f{1}{N}\sum_{k=1}^N\l_k^{-s}$; see \ref{app3} for the derivation. In addition, since the
minimum eigenvalue of the asymptotic distribution 
$\l_{\min}=1+\a-2\sqrt{\a}$ is approximately close to $+0$ when $\a\to+1$,
if $m$ eigenvalues are regarded as minimum eigenvalues, 
where $m\sim O(1)$, then by
using L'H$\hat{\rm o}$pital's rule, we can estimate the
asymptotic form of $q_w$ as follows:
\bea
q_w\eq\lim_{\l_{\min}\to+0}
\f{\f{m}{N}\l_{\min}^{-2}+\f{1}{N}\sum_{k=m+1}^N\l_k^{-2}}
{\left(\f{m}{N}\l_{\min}^{-1}+\f{1}{N}\sum_{k=m+1}^N\l_k^{-1}\right)^2
}\nn
\eq\lim_{\l_{\min}\to+0}\f{\f{2m}{N}\l_{\min}^{-1}}{
\f{2m}{N}\left(\f{m}{N}\l_{\min}^{-1}+\f{1}{N}\sum_{k=m+1}^N\l_k^{-1}\right)
}\nn
\eq\f{N}{m}.
\eea
Since $E[\l^{-2}]$ increases faster than $(E[\l^{-1}])^2$, 
$q_w$ increases. This is consistent 
with our finding that $q_w=\f{\a}{\a-1}$. Moreover, if $\a\gg1$,
then $1\le q_w\le(\l_{\max}/\l_{\min})^2$, and
\bea
\lim_{\a\to\infty}\left(\f{1+\a+2\sqrt{\a}}{1+\a-2\sqrt{\a}}\right)^2\eq1,
\eea
where the maximum asymptotic eigenvalues are 
$\l_{\max}=1+\a+2\sqrt{\a}$ and 
$\l_{\max}^{-s}\le 
E[\l^{-s}]\le\l_{\min}^{-s}$. This is also consistent with our 
finding that $q_w=\f{\a}{\a-1}$.
\subsection{Numerical simulation\label{sec4.4}}
Although we presented a theoretical discussion of the potential of an investment system, 
using the self-averaging property of the investment risk per asset $\ve$ and the
concentrated investment level $q_w$, 
we assumed that the replica number $n$ in \sref{eq40} was a real number 
 in Subsection \ref{sec4.2}. 
In the previous subsection, we presented some mathematical interpretations for 
our findings. However, it is not guaranteed mathematically that the replica number 
$n\in{\bf R}$ is applicable; we thus need to verify that we may use
this assumption in order to legitimize the findings based on our proposed method.
In this subsection, we perform a numerical simulation, and we then compare the results of our proposed method, the numerical results, and the results 
from the analytical operations research procedure. 

In this numerical simulation, the number of investment 
{outlets} was $N=10^3$, and the number of scenarios was $p\in[1200,8000]$; 
the scenario ratio was $\a\in[1.2,8.0]$. In addition, 
we assessed $J^{-1}=(XX^{\rm T})^{-1}$, the inverse of the variance-covariance matrix defined by the randomly assigned return rate 
matrix $X$; the return rates on assets were independently and 
identically distributed with a standard normal distribution. We then solved 
\sref{eq3} for the optimal portfolio in order 
to estimate the minimal investment 
risk per asset $\ve(X)$ and the
concentrated investment level $q_w(X)$.
Finally, we averaged them over $100$ sets of the 
return rate matrix.

\begin{figure}[tb]
\begin{center}
\includegraphics[width=0.8\hsize]{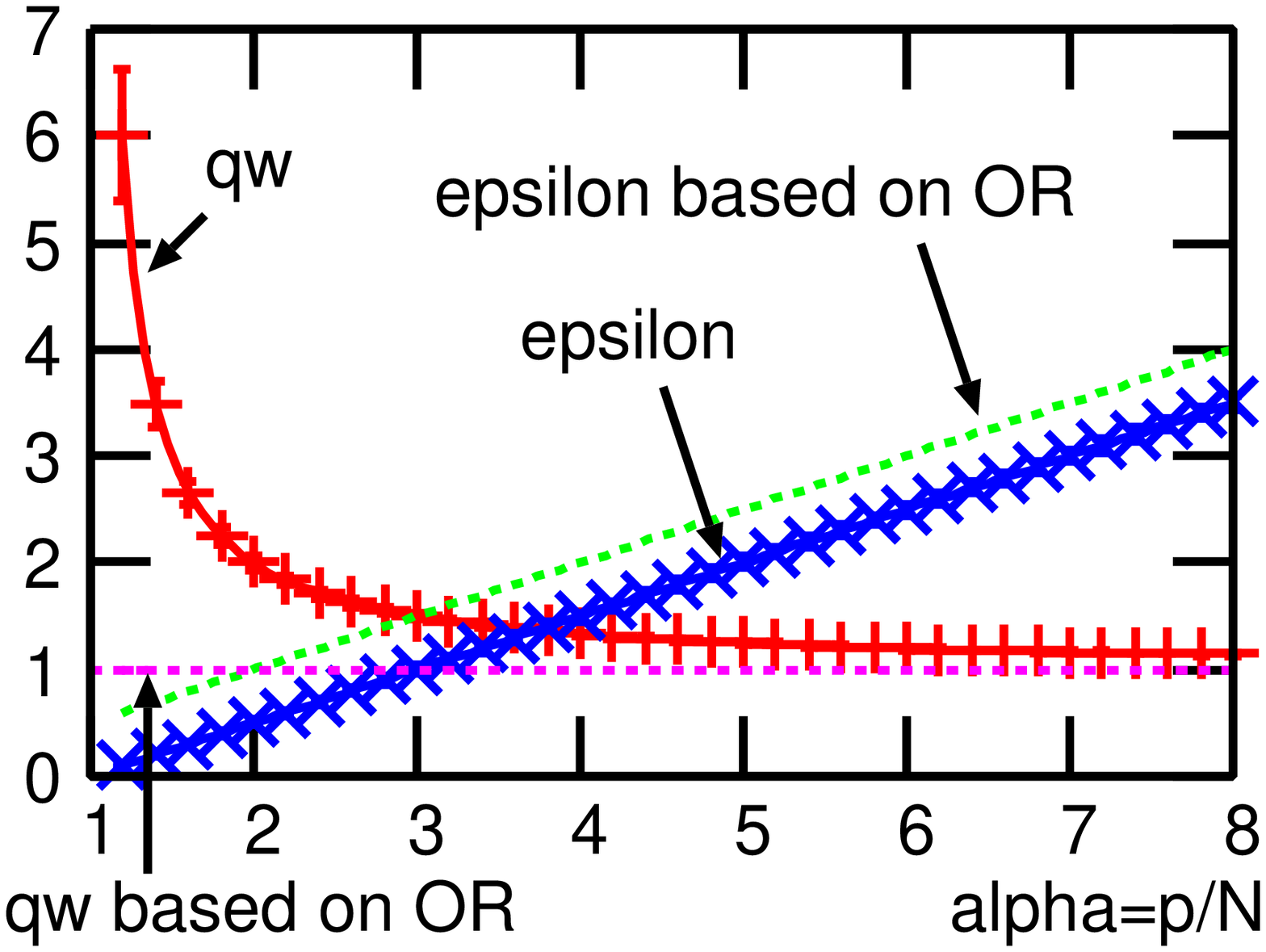}
\caption{\label{fig1}
The investment risk $\ve$ and the concentrated investment level $q_w$ are shown 
for the case in which the return rate $x_{k\mu}$ is independently and identically 
 distributed with a standard normal distribution.
The horizontal axis indicates the scenario ratio $\a=p/N$, and the vertical axis 
 shows the investment risk and the
concentrated investment level $q_w$.
The two solid lines (results obtained by our proposed approach)
 and the two dotted lines (results obtained by the operations research approach)
are theoretical results. The 
markers with error bars are the numerical results
evaluated using the optimal solution according to a return rate 
 which was randomly assigned.
In the simulation, the number of investment outlets $N$ was $10^3$,
and we averaged $100$ return rate matrices 
 $X=\left\{\f{x_{k\mu}}{\sqrt{N}}\right\}\in{\cal M}_{N\times p}$. 
This figure shows that the results obtained by our proposed approach (solid 
 lines) and the numerical results (markers with error bars) are in agreement. 
 On the other hand, the results obtained by the operations research approach 
 (dotted lines) do not coincide with the others. Thus, unfortunately, 
the approach based on maximizing the expected utility cannot 
propose an
optimal investment strategy.
}
\end{center}
\end{figure}

In Fig. \ref{fig1}, three minimal investment risks per asset and three concentrated investment levels 
are shown. The horizontal axis indicates the scenario ratio $\a=p/N$, and the
vertical axis shows the two indicators. The 
results of our proposed approach are indicated by solid lines, the numerical 
results are indicated by markers with error bars, and the results of the operations 
research approach are indicated by dotted lines. The results of our method (solid lines) and the numerical results (markers with error bars) are in agreement.
For this numerical simulation, we considered the case in which
we have a priori knowledge of the {return rates}. Thus, it turns out that our proposed approach 
can precisely assess the potential of an investment system. On the 
other hand, the dotted lines are based on a scenario in which the expected utility is maximized, and these results do not coincide with the others.
Unfortunately, this indicates 
that the approach {based on} maximizing the expected utility is unable to  
determine the 
optimal investment strategy and may instead
provide a misleading portfolio which is not guaranteed to be optimal 
with respect to particular set of return rates.

\section{Summary and future work}
In this paper, we analyzed 
the potential of an optimal solution to the mean-variance model, which is 
widely used for the portfolio 
optimization problem; in particular, we analyzed its potential 
investment {risk and
 the
concentrated 
investment level using self-averaging and replica analysis.}
We used the example of the rock-paper-scissors game with two subjects 
as a context for the optimization of stochastic phenomena. We noted that the
minimal expected investment risk (from our discussion of an annealed disorder 
system) is not always in agreement with
the expected minimal investment risk (from our discussion of a quenched 
disorder system).
We discussed whether 
the optimal investment strategy 
which was derived using the analytical procedure that is widely used in operations research and the
maximization of the expected utility based on an annealed disorder system 
are valid for use with actual return rates 
on assets. 
From the relationship in \sref{eq7}, based on the more general 
formulation, we determined 
that the minimal expected investment risk obtained by the operations research 
approach was not smaller than the expected minimal 
investment risk. 
However, it does not provide useful information for an investment strategy since it underestimates the expected minimal investment risk. 
The main reasons for this are as follows. (1) At the start of an investment, there is no a priori knowledge of the future 
return rates on the assets. (2) The computational complexity required to assess the inverse of the optimal 
solution matrix increases with the cube of the number of investment outlets. 
(3) In order to precisely assess the potential investment risk, 
it is necessary to average the minimal investment risk with the actual return 
rate. In order to solve these problems, we used {probabilistic 
reasoning} to reformulate the
portfolio optimization problem;
we also used the Chernoff inequality and replica analysis to determine a tighter upper bound for the cumulative distribution of the investment risk.
From an analytical result for the rate function that was 
derived from replica analysis, we clarified the
self-averaging property of the investment risk.
Thus, we determined that the
minimal investment risk for the case in which 
complete information on the return rates is known a priori is in agreement with 
the minimal investment risk for the case in which the return rate matrix is {averaged}. 
We are thus able to evaluate the potential investment risk in an actual investment system.
From this, we have solved the first and third problems that we listed {above}.
Furthermore, by using replica analysis, we estimated
two indicators for the optimal portfolio: the investment risk and the concentrated investment level; this was done without resolving the optimal portfolio directly, and this resolved the
second problem. 
We found that the concentrated investment level obtained by our proposed approach was 
consistent with the intuitively obvious choice for an optimal investment 
strategy; we considered cases in which 
the scenario ratio approached $1$ and in which the ratio was 
sufficiently large. 
We compared the results of our proposed method, the results obtained by the operations research approach, and the results obtained from a numerical simulation.
The results of our method were in agreement with the results of the numerical simulation,
but they did not coincide with the results of the operations research approach. As discussed above, 
although our findings are based on the mean-variance model with only a budget 
constraint and a return 
rate which is independently and identically 
distributed with a standard normal distribution, 
the relationship in \sref{eq7} and our findings imply that the approach based on maximizing the expected utility 
is not able to determine the most desirable strategy for actual investments.

In our future work, although 
for simplicity
we considered only a budget constraint in this paper, 
in order to make our method more realistic,  
we wish to determine the optimal solutions under other constraints, such as limits on expected 
gross earnings, short-selling restrictions, and upper {and lower} limits for each asset. In particular, we wish to consider whether
these problems can be resolved by using other analytical approaches of statistical mechanical informatics, such as the belief propagation method, 
random matrix integrals, or the Markov Chain Monte Carlo {method}. It is also necessary
to confirm the self-averaging property of the risk function for cases other than the mean-variance model, such as for 
the mean-absolute deviation model or the expected shortfall 
model (for {these models}, typical behaviours of investment risk were evaluated by 
Ciliberti and M$\acute{\rm e}$zard). In addition, 
in order to clarify the mathematical structure of this optimization 
problem, we assumed that the return rates on assets were independently and 
identically distributed with a standard normal distribution, however,
in an actual investment market, the return rate is not always independently and 
 identically distributed. We would thus like to quantify the effects of this correlation on the indicators.
Thus, although several models used in operations research have been 
proposed for assessing investment systems, 
in many cases, only the expected utility has been
maximized; that is, only the 
annealed disorder system has been 
analyzed. 
The
portfolio optimization problem is an undeveloped field, and many issues have not yet been considered.



The author thanks R. Wakai, Y. Shimazaki, and I. Kaku for their fruitful 
 discussions. The author is also grateful to Y. Takemoto, I. Arizono, and 
 T. Mizuno for valuable 
 comments. 
This paper is an improved and expanded version of a previous
paper by the same author\cite{Shinzato-RIMS}, 
 which was an unrefereed conference paper written in Japanese. 
This work was supported in part by a Grant-in-Aid for Young 
 Scientists (B), No. 24710169.
\appendix
\section{Properties of the rate function\label{app1}}
We introduce the properties of the rate function $R(\eta)$; these properties support the discussion in Section 
\ref{sec4}. In this appendix, for convenience, we define the function 
\bea
\phi(u)\eq\log E[e^{uY}].
\eea
Moreover, we will discuss only 
$Pr[\eta\le Y]\le e^{-R(\eta)}$ and 
$R(\eta)=\mathop{\max}_{u>0}\left\{u\eta-\phi(u)\right\}$, 
but we note that $Pr[\eta\ge Y]\le e^{-R(\eta)}$ and 
$R(\eta)=\mathop{\max}_{u<0}\left\{u\eta-\phi(u)\right\}$ can be verified 
in a similar way.
\subsection{$R(\eta)\ge0$}
For any $\eta\in{\bf R}$, $R(\eta)\ge0$ holds. Also,
for any $u>0$, $R(\eta)=\mathop{\max}_{u>0}\left\{u\eta-\phi(u)\right\}\ge 
u\eta-\phi(u)=R(\eta,u)$ in the limit as $u$ goes to $+0$,  that is, we obtain 
$\lim_{u\to+0}R(\eta,u)=0$ for 
$R(\eta)\ge0$. In addition, if $R(\eta)<0$, since 
$Pr[\eta\le Y]\le1< e^{-R(\eta)}$, 
$R(\eta)\ge0$ implies an intuitive upper bound for the 
cumulative distribution.

\subsection{When $E[Y]\ge\eta$, $R(\eta)=0$}
When $\eta$ is less than or equal to $E[Y]$, the expectation of $Y$, 
that is, $E[Y]\ge\eta$, then
$R(\eta)=0$. Since $e^{uY}$ is a convex function of $Y$, for any 
$u>0$, 
$\phi(u)\ge \log e^{uE[Y]}=uE[Y]$, and we obtain 
$0\ge uE[Y]-\phi(u)$.
If $\eta=E[Y]$, then we obtain $R(E[Y])=0$ from $R(E[Y],u)=uE[Y]-\phi(u)\le0$.
In addition, if $E[Y]\ge \eta$, then
$0\ge uE[Y]-\phi(u)\ge 
u\eta-\phi(u)=R(\eta,u)$, and we obtain $R(\eta)=0$. 
That is, this property may intuitively imply 
$E[Y]=\sup\left\{\eta|R(\eta)=0\right\}$.
\subsection{$R(\eta)$ is a convex function}
$R(\eta)$, which is derived from the Legendre transformation of a convex 
function, is a convex function of $\eta$.
Thus, {for any $\forall\l\in[0,1)$, $R(\eta)$ and $R(\xi)$,} 
\bea
\l R(\eta)+(1-\l)R(\xi)&\ge&
\l R(\eta,u)+(1-\l)R(\xi,u)\nn
\eq u(\l\eta+(1-\l)\xi)-\phi(u)\nn
\eq R(\l\eta+(1-\l)\xi,u),\label{eqa2}
\eea
where $u$ is nonnegative. Thus, the $\l 
R(\eta)+(1-\l)R(\xi)\ge R(\l\eta+(1-\l)\xi)$ 
is obtained by maximizing both sides of \sref{eqa2} for 
$u>0$.

\section{Calculation of replica analysis\label{app2}}
In this appendix, we analytically evaluate 
$E[Z^n(\b,X)]$ using replica analysis; however, in general, it is 
difficult to estimate 
$E[Z^n(\b,X)]$ for any $n\in{\bf R}$\cite{Nishimori}. Direct evaluation of 
$E[Z^n]$, the $n$-th moment of a nonnegative random variable $Z\ge0$, at any $n\in{\bf R}$, is 
not possible unless the random variable 
follows a log-normal distribution\cite{Ogure,Tanaka}.
In particular, it is not easy to assess the partition function $Z(\b,X)$, defined in integral form in \sref{eq17}, 
with a fixed return rate matrix $X$. 
If we could calculate this directly, 
we could easily solve \sref{eq25} and \sref{eq26} without needing to refer to the Helmholtz free energy;
however,  it is difficult to directly  evaluate a partition function with a
fixed return matrix in this model.
We can solve $E[Z^n(\b,X)]$ with the replica number {$n\in{\bf N}$} because it is
comparatively easy to calculate
$E[Z^n(\b,X)]$ for any replica number $n\in{\bf N}$, and this can be used to estimate 
$E[Z^n(\b,X)]$ at any replica number $n\in{\bf R}$. 
Intuitively, for instance, it is possible to expand $(a+b)^2=a^2+2ab+b^2$ and 
$(a+b)^3=a^3+3a^2b+3ab^2+b^3$ with finite terms, although 
it is not possible to obtain a finite expansion of $(a+b)^{2.5}$. 
Nevertheless, it is trivial that this expansion will be between the square and the cube of $(a+b)$, 
that is,  
$(a+b)^2<(a+b)^{2.5}<(a+b)^3$. 
Thus, as a first step, we can estimate $E[Z^n(\b,X)]$ at replica number $n\in{\bf N}$, 
and then use this to estimate $E[Z^n(\b,X)]$ at replica number $n\in{\bf R}$. This approach is
called replica analysis.

We can evaluate $E[Z^n(\b,X)]$ at $n\in{\bf N}$, as follows:
\bea
E[Z^n(\b,X)]\eq
E
\left[
\left(
\area d{\bf w}P_0({\bf w})
e^{-\b{\cal H}({\bf w}|X)}
\right)^n
\right]\nn
\eq
\area \prod_{a=1}^nd{\bf w}_aP_0({\bf w}_a)\nn
&&
E
\left[
\exp\left(-\f{\b}{2}\sum_{\mu=1}^p\sum_{a=1}^n
\left(\f{1}{\sqrt{N}}\sum_{i=1}^Nx_{i\mu}w_{ia}\right)^2
\right)
\right]\label{eqa3},
\eea
where ${\bf w}_a=(w_{1a},\cdots,w_{Na})^{\rm T}\in{\bf 
R}^N,(a=1,2,\cdots,n)$. Moreover, since $P_0({\bf w}_a)$ is the
coefficient used to average the return rate matrix $X$, 
it can be separated. 

We now introduce the Dirac delta function $\d(x)$ {in order to} use it to average the $x_{k\mu}$. 
The Dirac delta function $\d(x)$ is one of the most widely used generalized 
functions, defined
for any $f(x)$ as
\bea
f(w)\eq\area dvf(v)\d(v-w).\label{eqa37}
\eea
{This function returns the function $f(v)$ when the argument 
of $\d(v-w)$ on the right-hand side, $v-w$, is 0, that 
is $f(w)$. Thus, }
if a constant function is used in \sref{eqa37}, for example,  
$f(x)=1$, then 
\bea
\area dz\d(z)\eq1.
\eea
In addition, the Fourier transform of the Dirac delta function, 
\bea
\d(z)\eq
\f{1}{2\pi}\area
due^{iuz},
\eea
can be obtained if the imaginary unit $i=\sqrt{-1}$ is employed.
Thus, the integrand in \sref{eqa3} can be written as 
\bea
&&\exp\left(
-\f{\b}{2}
\left(\f{1}{\sqrt{N}}\sum_{i=1}^Nx_{i\mu}w_{ia}\right)^2
\right)\nn
\eq
\f{1}{2\pi}\area
dv_{\mu a}du_{\mu a}
\exp\left({-\f{\b v_{\mu a}^2}{2}}+iu_{\mu a}
\left(v_{\mu a}-\f{1}{\sqrt{N}}\sum_{i=1}^Nx_{i\mu}w_{ia}\right)
\right).\quad
\eea
Substituting this into \sref{eqa3}, we obtain
\bea
E[Z^n(\b,X)]
\eq
\f{1}{(2\pi)^{pn}}
\area
\prod_{a=1}^n
d{\bf w}_aP_0({\bf w}_a)
d{\bf u}_ad{\bf v}_a\nn
&&
\exp
\left(
i\sum_{\mu=1}^p\sum_{a=1}^nu_{\mu a}v_{\mu a}
-\f{\b}{2}
\sum_{\mu=1}^p\sum_{a=1}^n
v_{\mu a}^2
\right)
\nn
&&E_X\left[
\exp\left(
-\f{i}{\sqrt{N}}
\sum_{\mu=1}^p\sum_{a=1}^n\sum_{i=1}^N
u_{\mu a}x_{i\mu}w_{ia}
\right)
\right],
\eea
{where ${\bf u}_a=(u_{1a},\cdots,u_{pa})^{\rm 
T}\in{\bf R}^p$ and ${\bf v}_a=(v_{1a},\cdots,v_{pa})^{\rm T}\in{\bf R}^p$, $(a=1,2,\cdots,n)$}. Since the return rate $x_{i\mu}$ is independently and identically distributed with a
standard normal distribution, the expectation of $x_{i\mu}$ is 
\bea
E\left[
\exp\left(-\f{ix_{i\mu}}{\sqrt{N}}
\sum_{a=1}^nu_{\mu a}w_{ia}
\right)
\right]
\eq
\exp
\left(
-\f{1}{2N}
\left(\sum_{a=1}^nu_{\mu a}w_{ia}\right)^2
\right)
\eea
where {$
\area
\f{dx}{\sqrt{2\pi\s^2}}
e^{-\f{(x-m)^2}{2\s^2}+ix\theta}
=
e^{im\theta-\f{\s^2\theta^2}{2}}$.}
Thus we obtain
\bea
E[Z^n(\b,X)]
\eq
\f{1}{(2\pi)^{pn}}
\area
\prod_{a=1}^n
d{\bf w}_aP_0({\bf w}_a)
d{\bf u}_ad{\bf v}_a
\nn
&&\exp
\left(
i\sum_{\mu=1}^p\sum_{a=1}^nu_{\mu a}v_{\mu a}
-\f{\b}{2}
\sum_{\mu=1}^p\sum_{a=1}^n
v_{\mu a}^2
\right.\nn
&&
\left.
-\f{1}{2N}
\sum_{\mu=1}^p
\sum_{i=1}^N
\left(\sum_{a=1}^n
u_{\mu a}w_{ia}
\right)^2
\right).
\eea
We then substitute 
\bea
\label{eqa9}
q_{wab}\eq
\f{1}{N}
\sum_{i=1}^N
w_{ia}w_{ib}
\eea
and obtain
\bea
-\f{1}{2{N}}
\sum_{\mu=1}^p
\sum_{i=1}^N
\left(\sum_{a=1}^n
u_{\mu a}w_{ia}
\right)^2
\eq-\f{1}{2}
\sum_{\mu=1}^p
\sum_{a=1}^n
\sum_{b=1}^n
u_{\mu a}
u_{\mu b}q_{wab}.
\eea
From this technique, we obtain
\bea
E[Z^n(\b,X)]
\eq\label{eqa11}
\mathop{\rm Extr}_{Q_w,\tilde{Q}_w}
\left\{\f{1}{(2\pi)^{pn}}
\area
\prod_{a=1}^n
d{\bf w}_aP_0({\bf w}_a)
d{\bf u}_ad{\bf v}_a
\right.\nn
&&\left.
\exp
\left(
i\sum_{\mu=1}^p\sum_{a=1}^nu_{\mu a}v_{\mu a}
-\f{\b}{2}\sum_{\mu=1}^p\sum_{a=1}^nv_{\mu a}^2
-\f{1}{2}
\sum_{\mu=1}^p
\sum_{a,b}
u_{\mu a}u_{\mu b}q_{wab}
\right.
\right.
\nn
&&
\left.
\left.
-\f{1}{2}
\sum_{a,b}\tilde{q}_{wab}
\left(\sum_{i=1}^Nw_{ia}w_{ib}-Nq_{wab}\right)
\right)
\right\},
\eea
where $\sum_{a,b}$ means 
$\sum_{a=1}^n\sum_{b=1}^n$ and $\mathop{\rm Extr}_Af(A)$ 
are the extrema of $f(A)$ with respect to $A$. In 
order to satisfy the constraint in \sref{eqa9}, we use the 
auxiliary variable $\tilde{q}_{wab}$. {Moreover 
$Q_w=\left\{q_{wab}\right\}\in{\cal M}_{n\times n}$ and 
$\tilde{Q}_w=\left\{\tilde{q}_{wab}\right\}\in{\cal M}_{n\times n}$ are the order 
parameter matrices.}

We can separate the integral of $u_{\mu a},v_{\mu a}$ from the integral of 
$w_{ka}$. We evaluate the integral of $u_{\mu a},v_{\mu a}$, 
\bea
&&
\f{1}{(2\pi)^{pn}}\area 
\prod_{\mu=1}^p\prod_{a=1}^n
du_{\mu a}dv_{\mu a}\nn
&&
\exp\left(
\sum_{\mu=1}^p
\left(
i\sum_{a=1}^nu_{\mu a}v_{\mu a}
-\f{\b}{2}\sum_{a=1}^nv_{\mu a}^2
-\f{1}{2}\sum_{a,b}u_{\mu a}u_{\mu b}q_{wab}
\right)
\right)\nn
\eq\left\{
\f{1}{(2\pi)^n}\area d{\bf u}d{\bf v}e^{i{\bf u}^{\rm T}{\bf 
v}-\f{\b}{2}{\bf v}^{\rm T}{\bf v}-\f{1}{2}{\bf u}^{\rm T}Q_w{\bf u}}
\right\}^p\nn
\eq\exp\left[
-\f{p}{2}\log\det\left|I+\b Q_w\right|
\right],
\eea
{
where $I\in{\cal M}_{n\times n}$ is the 
identity matrix. Because this} is independent of the
scenario index $\mu$, 
we can estimate the integral using two novel vectors, 
${\bf u}=(u_1,\cdots,u_n)^{\rm T}\in{\bf R}^n$, and ${\bf 
v}=(v_1,\cdots,v_n)^{\rm T}\in{\bf R}^n$. On the other hand, we can calculate the integral of $w_{ka}$,
\bea
&&\area \prod_{k=1}^N\prod_{a=1}^ndw_{ka}P_0({\bf w}_a)
\exp\left(-\f{1}{2}\sum_{k=1}^N\sum_{a,b}\tilde{q}_{wab}w_{ka}w_{kb}+\f{N}{2}{\rm Tr}Q_w\tilde{Q}_w\right)\nn
\eq\mathop{\rm Extr}_{{\bf k}}
\f{1}{(2\pi)^{\f{Nn}{2}}}\area \prod_{k=1}^N\prod_{a=1}^ndw_{ka}\nn
&&
\exp\left(\sum_{a=1}^nk_a
\left(\sum_{k=1}^Nw_{ka}-N\right)
-\f{1}{2}\sum_{k=1}^N\sum_{a,b}\tilde{q}_{wab}w_{ka}w_{kb}+\f{N}{2}{\rm Tr}Q_w\tilde{Q}_w\right)\nn
\eq
\mathop{\rm Extr}_{\bf k}
\exp\left[{-N{\bf k}^{\rm T}{\bf e}+\f{N}{2}{\rm Tr}Q_w\tilde{Q}_w}\right]
\left\{
\f{1}{(2\pi)^{\f{n}{2}}}\area d{\bf w}
e^{-\f{1}{2}{\bf w}^{\rm T}\tilde{Q}{\bf w}+{\bf k}^{\rm T}{\bf w}}
\right\}^N\nn
\eq
\mathop{\rm Extr}_{\bf k}
\exp\left[{-N{\bf k}^{\rm T}{\bf e}+\f{N}{2}{\rm Tr}Q_w\tilde{Q}_w}
-\f{N}{2}\log\det\left|\tilde{Q}_w\right|+\f{N}{2}{\bf k}^{\rm 
T}\tilde{Q}_w^{-1}{\bf k}
\right],\label{eqa13}
\eea
{
where ${\bf k}=(k_1,\cdots,k_n)^{\rm T}\in{\bf R}^n,{\bf 
e}=(1,\cdots,1)^{\rm T}\in{\bf 
R}^n$,}
${\bf w}$ is the prior probability of the portfolio, $P_0({\bf 
w}_a)$ is replaced by $\mathop{\rm 
Extr}_{k_a}\exp\left(k_a\left(\sum_{k=1}^Nw_{ka}-N\right)-\f{N}{2}\log2\pi\right)$, and because this is not dependent on the asset index $k$,
we can solve the integral using a novel vector
${\bf w}=(w_1,\cdots,w_n)^{\rm T}\in{\bf R}^n$.

We summarize this and rewrite the limit of $\f{1}{N}\log E[Z^n(\b,X)]$ 
for the number of investment outlets $N$ as $\Phi(n)$:
\bea
\Phi(n)\eq\lim_{N\to\infty}\f{1}{N}\log E[Z^n(\b,X)]\nn
\eq
\mathop{\rm Extr}_{{\bf k},Q_w,\tilde{Q}_w}
\left\{
-\f{\a}{2}\log\det\left|I+\b Q_w\right|
+\f{1}{2}{\rm Tr}Q_w\tilde{Q}_w
-\f{1}{2}\log\det\left|\tilde{Q}_w\right|
\right.\nn
&&\left.
-{\bf k}^{\rm T}{\bf 
e}+\f{1}{2}{\bf k}^{\rm T}\tilde{Q}_w^{-1}{\bf k}
\right\}.\qquad\qquad
\eea
Although a
sufficiently large number of investment outlets $N$ 
is required to guarantee 
that 
evaluating by using the order parameters 
$k_a,q_{wab},\tilde{q}_{wab}$ as defined in \sref{eqa11} and \sref{eqa13} 
is consistent with the constraints of \sref{eqa9} and \sref{eq2} in the replica analysis, 
our target indicator $\ve$ represents the minimal investment risk 
per asset; that is, since this is independent of the system 
size $N$, 
there will not be problems in the limit as $N$ approaches infinity.
It is preferable to normalize the investment risk per asset with respect to different sizes of investment markets, 
and this allows the comparison of potential investment risks.

We note two important points. First, 
in Subsection \ref{sec4.2}, we already mentioned that $\ve(X)=E[\ve(X)]$, since the investment risk is self-averaging.  From the
above discussion, we have verified that
\bea
E[\ve(X)]\eq-\lim_{\b\to\infty}\pp{}{\b}
\left\{
\lim_{N\to\infty}\f{1}{N}E[\log Z(\b,X)]
\right\}\nn
\eq-\lim_{\b\to\infty}\pp{}{\b}\left\{\lim_{n\to0}\pp{\Phi(n)}{n}\right\}\nn
\eq-\lim_{\b\to\infty}\pp{}{\b}\left\{
-\f{\a}{2}\log\f{\a}{\a-1}-\f{\b(\a-1)}{2}+\f{1}{2}-\f{1}{2}\log\b(\a-1)
\right\}\nn
\eq\f{\a-1}{2},
\eea
where we assume that the replica number $n$ is a
continuous number, and we use {the replica trick} $E[\log 
Z]=\lim_{n\to0}\pp{}{n}\log E[Z^n]$\cite{Nishimori,Ogure,Tanaka}. 
This result is consistent with that of \sref{eq52}.
Second, from the definition in \sref{eqa9}, since $q_{waa}$ is consistent with the 
concentrated 
investment level $q_{w}$ in \sref{eq5}, 
$q_w=q_{waa}=\f{1}{\b(\a-1)}+\f{\a}{\a-1}$.  However, 
since this is an optimal solution with a sufficiently 
large $\b$, 
\bea
q_{w}\eq\f{\a}{\a-1}.
\eea
Although we used replica analysis to analyze the minimal investment risk, 
we can also obtain the concentrated investment level of the optimal portfolio.
Fortunately, in the limit of very large $N$,
$q_w$ is finite; 
thus there is an advantage of using \sref{eq2} as the budget constraint.

\section{Random matrix approach for minimal investment risk and 
 concentrated investment level\label{app3}}
We show here that it is also possible to evaluate the two 
indicators, $\ve$ and $q_w$, by using an
asymptotic eigenvalue distribution of a random matrix\cite{Wakai}. As in the above discussion, 
we will consider only the case $\a=p/N>1$ in order to uniquely determine the optimal 
solution of \sref{eq1}.
Using the optimal solution defined in \sref{eq3}, 
from \sref{eq13} and \sref{eq14}, the
minimal investment risk per asset $\ve$ and the concentrated investment 
level $q_w$ are replaced, as follows: 
\bea
\ve
\eq\f{1}{2
\left(\f{1}{N}{\bf e}^{\rm T}J^{-1}{\bf e}\right)
},\\
q_w
\eq\f{\left(\f{1}{N}{\bf e}^{\rm T}J^{-2}{\bf e}\right)}{\left(\f{1}{N}{\bf e}^{\rm T}J^{-1}{\bf e}\right)^2}.
\eea
If $N$ is sufficiently large, we have 
\bea
\ve\eq\f{1}{2g(1)},\\
q_w\eq\f{g(2)}{(g(1))^2},
\eea
where 
$g(s)=\lim_{N\to\infty}
\f{1}{N}{\bf e}^{\rm T}J^{-s}{\bf e}
$. If we could analyze $g(s)$, then $\ve$ and $q_w$ could be precisely determined. 
It turns out that it is easy to assess $g(s)$ by using a random matrix ensemble.

For this ensemble of random matrices, 
we require the following two properties: 
(1) when the random matrix $X=\left\{\f{x_{k\mu}}{\sqrt{N}}\right\}\in{\cal M}_{N\times 
p}$ is decomposed as $X=UDV$, where $U\in{\cal M}_{N\times 
N}$ and 
$V\in{\cal M}_{p\times p}$ are orthogonal matrices and $D\in{\cal 
M}_{N\times p}$ is a diagonal rectangular matrix, then $U$ and $V$ are 
independently distributed with a Haar measure; 
(2) when $N$ is sufficiently large, the distribution of the
eigenvalues of the variance-covariance matrix $J=XX^{\rm T}$, 
for any return rate matrix $X$, is asymptotically close to 
$\rho(\l)=\lim_{N\to\infty}\sum_{k=1}^N\d(\l-\l_k)$, 
where $\l_k$ is the $k$th diagonal of $DD^{\rm T}={\rm diag}\left\{\l_1,\l_2,\cdots,\l_N\right\}\in{\cal 
M}_{N\times N}$; if $N$ and $p$ simultaneously approach infinity, then it is 
required that $\a=p/N\sim O(1)$. If these two properties 
are satisfied, then 
\bea
g(s)\eq\int_0^\infty d\l\rho(\l)\l^{-s}.
\eea
Moreover, if the return rates are independently and 
identically distributed with a standard normal distribution, 
then the random matrix $X$ satisfies the requirements for the random matrix ensemble 
described above\cite{Shinzato-Kabashima,Wakai}.

Next, we consider the asymptotic eigenvalue distribution. If the return rate $x_{k\mu}$ is independently and identically distributed, 
its mean and variance are respectively $0$ and $1$, and
the higher-order moments are finite, that is, {
$\left|E[(x_{k\mu})^s]\right|<\infty,(s=3,4,\cdots)$,} then the distribution of the
eigenvalues of the variance-covariance matrix $J=XX^{\rm T}$ of the
return rate matrix $X=\left\{\f{x_{k\mu}}{\sqrt{N}}\right\}\in{\cal 
M}_{N\times p}$ is asymptotically close to 
\bea
\rho(\l)\eq
\left[1-\a\right]^+\d(\l)+\f{
\sqrt{
\left[\l-\l_-\right]^+
\left[\l_+-\l\right]^+
}
}{2\pi\l},
\eea
where $\d(u)$ is the Dirac delta function, $[u]^+=\max(0,u)$, and 
$\l_\pm=1+\a\pm2\sqrt{\a}$\cite{Bai, MP,Tulino}.
This eigenvalue distribution $\rho(\l)$ is called the Mar$\check{\rm 
c}$enko-Pastur law, and 
this distribution can be regarded as the limit distribution for the eigenvalues, 
similar to the limit distribution (normal distribution) guaranteed by the central limit theorem.

The eigenvalues in this distribution can be easily calculated: 
\bea
g(1)
\eq\f{\l_++\l_-}{4\sqrt{\l_+\l_-}}-\f{1}{2}\nn
\eq\f{1}{\a-1}\\
g(2)\eq\f{\sqrt{\l_+\l_-}}{4}
\left(\f{\f{1}{\l_-}-\f{1}{\l_+}}{2}\right)^2\nn
\eq\f{\a}{(\a-1)^3},
\eea
where 
\bea
\int \f{dx}{\sqrt{ax^2+bx+c}}\eq-\f{1}{\sqrt{|a|}}\sin^{-1}\f{2ax+b}{\sqrt{b^2-4ac}},\qquad(a<0)\\
\int \f{dx}{x\sqrt{ax^2+bx+c}}\eq
\f{1}{\sqrt{|c|}}\sin^{-1}\f{2c+bx}{x\sqrt{b^2-4ac}},\qquad(c<0).
\eea
Thus, 
\bea
\ve\eq\f{\a-1}{2}\\
q_w\eq\f{\a}{\a-1}.
\eea
This result is consistent with 
the result we obtained by replica analysis and numerical simulation.

\end{document}